# LATENT CLASS ANALYSIS FOR RELIABLE MEASURE OF INFLATION EXPECTATION IN THE INDIAN PUBLIC

Sunil Kumar
Alliance University, Bangalore, India

## ABSTRACT

The main aim of this paper is to inspect the properties of survey based on household's inflation expectations, conducted by Reserve Bank of India (RBI). It is theorized that the respondent's answers are exaggerated by extreme response bias. Latent class analysis (LCA) has been hailed as a promising technique for studying measurement errors in surveys, because the model produces estimates of the error rates associated with a given question of the questionnaire. I have identified a model with optimum performance and hence categorize the objective as well as reliable classifiers or otherwise.

**Keywords:** Inflation Expectation, Latent Class Analysis, Measurement error, classification.

## 1. Introduction

In household inflation expectation surveys, we have seen a robust increase in cross sectional dispersion during the recession periods which is at odds with standard rational expectation models that are at the center of most of the forecasting models of central banks. In recent years, research has moved away from full information rational expectations models towards a framework in which some form of limited information or bounded rationality is assumed. These models leave no room for heterogeneity or disagreement.

According to the RBI's questionnaire on Inflation expectation survey of household's which comprising of the future expectations about price questions i.e. Block 2 &3: Expectations of respondents on prices in next 3 months and a year ahead. Especially important for the conduct of monetary policy are the inflation expectations of households. Although there are numerous studies on the performance of survey based inflationary expectations (e.g. Ang et al. 2007, Scheufele 2010), there is a constant debate on applicability of households' inflation expectations to forecast changes in the inflation dynamics. It was confirmed in various papers that the inflation forecasts provided by professionals are very useful in predicting changes in the price level. Nevertheless, it was also shown that the households' inflation expectations provide better forecasts than time-series models or models based on the Phillips curve (Ang et al. 2007, p. 1191).

The respondents covered in the survey may incorporate extreme responses to complain against the government. Extreme response bias refers to the tendency to either uniformly endorse an optimistic view about the economy (positive bias) or uniformly report a negative perception about the economy (negative bias), irrespective of the information content sought (see Baumgartner and Steenkamp, 2006).

So, in this paper, I particularly address the issue of identifying extreme response bias using data from the Inflation expectations survey of households, during the ($3^{rd}$ and $4^{th}$) quarters of 2011 and first quarter of 2012 (i.e. $25^{th}$, $26^{th}$ and $27^{th}$ round data). The present study is only limited to three rounds of data due to the non-availability of the all-round data.

The paper is arranged in the following sections. In section 2, the data of Inflation expectations survey is described. In section 3, I discussed how Latent class models may be used to handle the extreme response bias in the Inflation expectations survey. Section 4 discusses the latent class model with all possible indicators, applied to the measurement of extreme response bias. Section 5 concludes with a discussion on results.

## 2. Inflation Expectations Survey of Households – (25 to 27) round data

**Description of the data**

Reserve Bank of India has been conducting Inflation Expectations Survey of Households (IESH) on a quarterly basis, since September 2005. The survey elicits qualitative and quantitative responses for three-month ahead and one-year ahead period on expected price changes and inflation. Inflation expectations of households are subjective assessments and are based on their individual consumption baskets and therefore may be different from the official inflation numbers released periodically by the government. Again, they may not be treated as forecast of any official measure of inflation, though these inflation expectations provide useful inputs on directional movements of future inflation. But in the present study, only three quarters data i.e. July-Sept, 2011; Oct-Dec, 2011 and Jan – March, 2012, is available from the RBI, respectively. A total of 11,793 interviews were used, where 100% response from $25^{th}$ and $26^{th}$ round and 94.8 % is the response rate from $27^{th}$ round is considered in the analysis because of unit and item non response.

**Sampling Design and Data collection**

The survey is conducted simultaneously in 12 cities that cover adult respondents of 18 years and above. The major metropolitan cities, viz., Delhi, Kolkata, Mumbai and Chennai are represented by 500 households each, while another eight cities, viz., Jaipur, Lucknow, Bhopal, Ahmedabad, Patna, Guwahati, Bengaluru and Hyderabad are represented by 250 households each. The respondents having a view on perceived current inflation are well spread across the cities to provide a good geographical coverage. The male and female respondents in the group are approximately in the ratio of 3:2. The survey schedule is organized into seven blocks covering the respondent profile (block 1), general and product-wise price expectations (block 2 and 3), feedback on RBI's action to control inflation (block 4), current and expected inflation rate (block 5), amount paid for the purchase of major food items during last one month (block 6) and the expectations on changes in income/wages (block 7).

The response options for price changes are (i) price increase more than current rate, (ii) price increase similar to current rate, (iii) price increase less than current rate, (iv) no change in prices, and (v) decline in prices. The inflation rates are collected in intervals - the lowest being 'less than 1 per cent' and the highest being '16 per cent and above' with 100 basis point size for all intermediate classes.

The descriptive statistics of the analyzed data is given in Table 1 and Table 2.

Table 1: Percentage of respondent's product wise expectations

| Round No./Survey period (Quarter ended) | 25 July-Sept, 2011 | 26 Oct-Dec, 2011 | 27 Jan-Mar, 2012 | 25 July-Sept, 2011 | 26 Oct-Dec, 2011 | 27 Jan-Mar, 2012 |
|---|---|---|---|---|---|---|
| | Three-month ahead (% age of respondents) | | | One year ahead (% age of respondents) | | |
| Prices will increase | 97.3 | 96.1 | 98.2 | 96.0 | 97.1 | 98.3 |
| Prices increase more than current rate | 75.8 | 73.4 | 75.6 | 73.5 | 76.9 | 78.8 |
| Price increase similar to current rate | 15.4 | 13.2 | 15.9 | 16.7 | 12.6 | 14.0 |
| Price increase less than the current rate | 6.0 | 9.6 | 6.7 | 5.9 | 7.6 | 5.5 |
| No change in prices | 2.2 | 3.0 | 1.6 | 3.3 | 2.5 | 1.6 |
| Decline in prices | 0.6 | 0.9 | 0.2 | 0.7 | 0.4 | 0.1 |

Table 2: Percentage of respondents expecting general price movements in coherence with movements in price expectations of various products groups: three months ahead (one year ahead).

| Round No. | Survey Quarter | Food | Non-Food | Households durables | Housing | Cost of services |
|---|---|---|---|---|---|---|
| 24 | July-Sept, 2011 | 88.8 (92.4) | 86.2 (87.2) | 68.0 (71.1) | 84.4 (85.7) | 85.2 (86.3) |
| 25 | Oct-Dec, 2011 | 88.5 (92.6) | 83.1 (84.8) | 69.6 (69.7) | 74.8 (78.4) | 74.6 (80.9) |
| 26 | Jan-Mar, 2012 | 87.7 (91.8) | 82.7 (84.0) | 65.4 (65.7) | 84.1 (83.9) | 83.7 (85.9) |

## 3. Methodology of Latent Class Analysis

According to Steenkamp and Baumgartner (2006), the most promising way of accounting for extreme response bias is inclusion of statistical techniques in the data analysis. Thus, the technique that enables to account for extreme response style is the latent class analysis (LCA). LCA is a powerful method for estimating one or more parameters of a survey error model. These estimates can be used inter alia to evaluate the Mean Square Error (MSE) of an estimate and the error probabilities associated with a survey question, {see Biemer (2011)}.

According to Bandeen-Roche et al. (1997), one can measure an unobservable variable which is conceptually defined but cannot be measured directly through several categorical indicators. Application of latent class models is appropriate when the data to be analyzed are categorical (either nominal or ordinal categories). In this approach, relationships between discrete indicator variables (questions) and the latent variable are modeled. A characteristic feature of the latent class analysis is that the latent variables are also discrete. Thus, LCA may be viewed as a method for analyzing categorical data analogues to factor analysis that allows one to analyze the structure of relationships among manifest i.e. observed discrete variables, in order to characterize a categorical latent variable. In LCA the measurement errors in the observations are referred to collectively as classification errors or, simply, misclassification.

The basic assumption of latent class analysis is that the observed (manifest) variables are caused by a latent categorical variable — they are indicators of this unobservable (latent) variable. Relationships between these manifest variables can be either causal (an independent variable causes a dependent variable) or symmetrical. It is also assumed that the manifest variables are dependent on the latent variable and the manifest variables define the latent variable as well. Hence, within each class of the latter variable, the co-variation between manifest variables should not be higher than random covariation which is the local independence condition of latent class analysis {see Bertrand and Hafner (2011)}.

Here, in general, we consider $J$ - polytomous categorical variables (manifest variables), each of which contains $K_j$ possible outcomes, for ith households $i(= 1,2, \dots, n)$. The latent class model approximates the observed joint distribution of the manifest variables as the weighted sum of a finite number R, of constituent cross-classification tables. Let $\pi_{jrk}$ denote the cross-conditional probability that an observation in class $r = 1, 2, \dots, R$ produces the $k^{th}$ outcome on the $j^{th}$ variable with $\sum_{k=1}^{K_j} \pi_{jrk} = 1$. Let $p_r$ be the prior probabilities of latent class membership, as they represent the unconditional probability that an individual will belong to each class before taking into account the responses $Y_{ijk}$ provided on the manifest variables. The probability that an individual $i$ in class $r$ produces a particular set of $J$ outcomes on the manifest variables, assuming conditional independence of the outcomes $Y$ given class membership, is the product

$$f(Y_i; \pi_r) = \prod_{j=1}^{J} \prod_{k=1}^{K_j} (\pi_{jrk})^{Y_{ijk}}, \qquad (1)$$

The probability density function across all classes is the weighted sum

$$f(Y_i|\pi, p) = \sum_{r=1}^{R} f(Y_i; \pi_r) = \sum_{r=1}^{R} p_r \prod_{j=1}^{J} \prod_{k=1}^{K_j} (\pi_{jrk})^{Y_{ijk}}, \qquad (2)$$

The parameters $p_r$ and $\pi_{jrk}$ are estimated by the latent class model.

Given estimates $\hat{p}_r$ and $\hat{\pi}_{jrk}$ of $p_r$ and $\pi_{jrk}$, respectively, the posterior probability that each individual belongs to each class, conditional on the observed values of the manifest variables, are calculated by

$$\hat{P}(r_i|Y_i) = \frac{\hat{p}_r f(Y_i; \hat{\pi}_r)}{\sum_{q=1}^{R} \hat{p}_q f(Y_i; \hat{\pi}_q)}, \qquad (3)$$

where $r_i \in (1, 2, \dots, R)$.

It is important that the condition $R \sum_j (K_j - 1) + (R - 1) \leq n$ on the number of parameters should hold. Also, $R \sum_j (K_j - 1) + (R - 1) \leq (3^{12} - 1)$ i.e. one fewer than the total number of cells in the cross-classification table of the manifest variables, as then the latent class model will be unidentified.

With the help of poLCA, one of the statistical packages in $R$ environment, one can estimate the latent class model by maximizing the log-likelihood function

$$\ln L = \sum_{i=1}^{n} \ln \sum_{r=1}^{R} p_r \prod_{j=1}^{J} \prod_{k=1}^{K_j} (\pi_{jrk})^{Y_{ijk}}, \qquad (4)$$

with respect to $p_r$ and $\pi_{jrk}$, using the expectation-maximization (EM) algorithm (Dempster, Laird and Rubin (1977), McCutcheon (1987), McLachlan and Krishnan (1997), McLachlan and Peel (2000), Everitt and Hand (1981), Everitt (1984) and Linzer and Lewis (2011)). In EM algorithm, poLCA begin with arbitrary initial values of $\hat{p}_r$ and $\hat{\pi}_{jrk}$, and denote them $\hat{p}_r^0$ and $\hat{\pi}_{jrk}^0$. In the expectation step, calculate the missing class membership probabilities using equation (3), substituting in $\hat{p}_r^0$ and $\hat{\pi}_{jrk}^0$. In the maximization step, update the parameter estimates by maximizing the log likelihood function given these posterior $\hat{P}(r_i|Y_i)$, with $\hat{p}_r^{new} = \frac{1}{N}\sum_{i=1}^{N}\hat{P}(r_i|Y_i)$ as the new prior probability and $\hat{\pi}_{jr}^{new} = \frac{\sum_{i=1}^{N} Y_{ij}\hat{P}(r_i|Y_i)}{\sum_{i=1}^{N}\hat{P}(r_i|Y_i)}$ as the new class conditional outcome probabilities; $\hat{\pi}_{jr}^{new}$ is the vector of length $K_j$ of class-$r$ conditional outcome probabilities for the $j^{th}$ manifest variable; and $Y_{ij}$ is the $N \times K_j$ matrix of observed outcome $Y_{ijk}$ on that variable. The algorithm repeats these steps several times until the overall log-likelihood reaches a local maximum and further increments are less than some arbitrarily small value.

Lin and Dayton (1997) suggested three criteria for selecting the best model. Model should be identifiable by Bandeen-Roche et al. (1997) methodology. A sufficient condition for such local identifiably of the standard latent class model is suggested by Goodman (1974). Then, the likelihood-ratio Chi-square ($L^2$) statistic is used as a standard measure of discrepancy between observed and expected frequencies in the model. However, the likelihood ratio Chi-square test, although extensively used in statistical literature, has a number of important limitations. The major one is its limited use when dealing with sparse tables. The likelihood-ratio statistic does not provide enough control for the number of parameters in a model that can sometimes be very large even for models of modest size (McCutcheon (2002)).

These limitations are controlled by the recent development and use of several information criteria, such as the Akaike information criterion (AIC) (Akaike (1973)) and Bayesian information criterion (BIC) (Schwartz (1978)), each of which is designed to penalize models with larger numbers of parameters. AIC and BIC on the number of parameters in the model:

$$AIC = L^2 - 2*d.f. \qquad (5)$$

and

$$BIC = L^2 - d.f.*\{ln(n)\}, \qquad (6)$$

where $n$ is the sample size. Thus, models with lower values of information criteria have a better fit to a data, for a given number of parameters. When sample size is large, BIC is preferred fit statistics and for small to medium sample sizes, the AIC statistic is most commonly used.

Here, I will use the same set of assumptions used by Sinclair and Gastwirth (1996) and by Biemer (2004). I fit the LCA models on the question on expectations of respondents on prices in next 3 months and next one year in general ($D^1$ & J), food products (E & K), non-food products (F & L), household durables (G & M), housing (H & N) and services (I & O), with polytomous options of responses (i.e. price increase more than current rate; price increase similar to current rate; price increase less than current rate; no change in prices; decline in prices), based on the questionnaire of inflation expectations survey of households, by Department of Statistics and Information Management, Reserve Bank of India. After identifying the optimal number of response classes for the latent variable; I have identified indicator(s) for which responses are inconsistent.

## 4. Choosing the number of latent classes

To identify the classes of latent variable, poLCA package in R statistical computing environment has been used. Model with no covariates has been considered to identify the latent classes among respondents based on response (indicator) variables. Table 3 summarizes the statistic which supports the existence of heterogeneity among respondents. Table 3 depicts that the minimum BIC statistic is associated with 5 latent classes which corresponds that model with 5 classes is optimal.

Table 3: Parameters on converged latent class models without covariates.

| Number of Classes | p | LL | AIC | BIC |
|---|---|---|---|---|
| 2 | 98 | -112775.4 | 225746.8 | 226469.6 |
| 3 | 148 | -106079.1 | 212454.2 | 213545.7 |
| 4 | 198 | -102390 | 205176.0 | 206636.2 |
| 5 | 248 | **-98635.34** | **197766.7** | **199595.7** |
| 6 | 298 | -99649.35 | 199894.7 | 202092.5 |

---

[1] D, E, F, G, H, I, J, K, L, M, N, O - are the notations of the questions of questionnaire of the IESH survey, which are explained in Appendix - I

The predicted response probabilities indicate the differences in response among classes shown in figure 1.

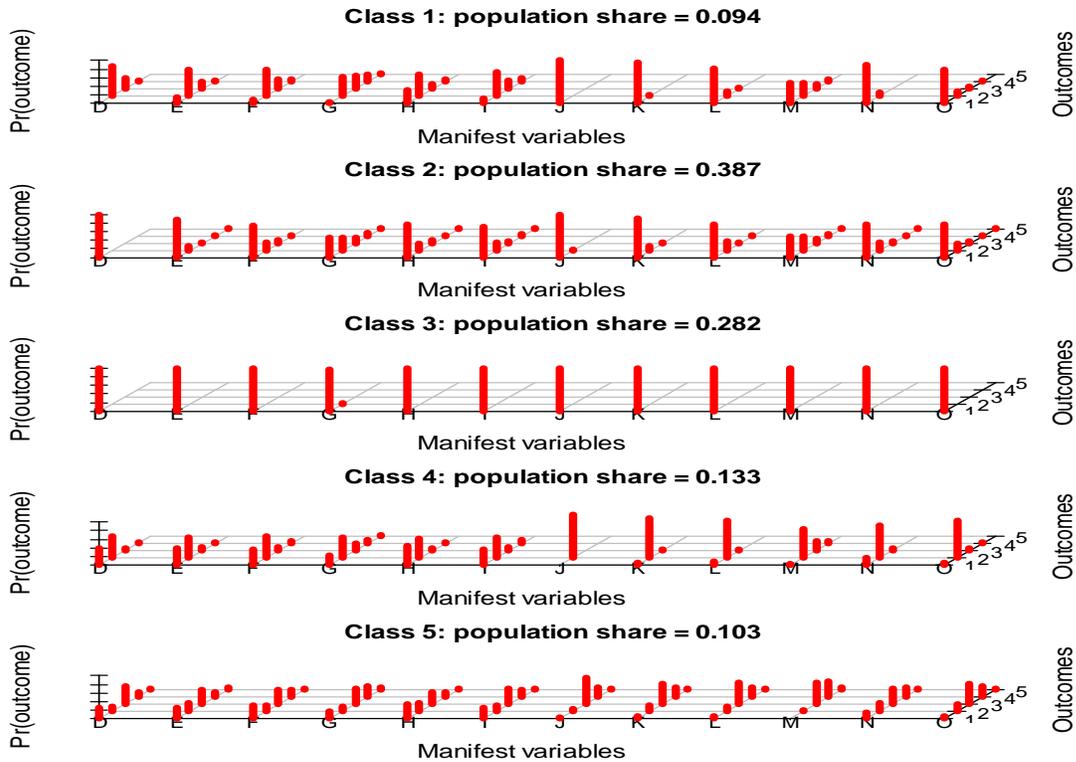

Figure 1

## 4.1 Characterizing classes and responses

The estimated class conditional probabilities were used to characterize the classes. Bold numbers in Table 4 classify the probabilities greater than 0.70 which are used to describe the identified classes.

Table 4: Estimated probabilities of answer for each question and each class

| Indicators | | Class 1 | Class 2 | Class 3 | Class 4 | Class 5 | Class 1 | Class 2 | Class 3 | Class 4 | Class 5 |
|---|---|---|---|---|---|---|---|---|---|---|---|
| | | | | 3 month ahead | | | | | One year ahead | | |
| General (D &J) | i | 0 | **0.998** | **1.000** | 0.398 | 0.270** | **1.000** | 0.992 | **1.000** | 0.000 | 0.030** |
| | ii | **0.711** | 0.002 | 0.000 | 0.502 | 0.118 | 0.000 | 0.007 | 0.000 | **1.000** | 0.073 |
| | iii | 0.247 | 0.000 | 0.000 | 0.063 | 0.425 | 0.000 | 0.001 | 0.000 | 0.000 | 0.619 |
| | iv | 0.040 | 0.000 | 0.000 | 0.037 | 0.135 | 0.000 | 0.000 | 0.000 | 0.000 | 0.239 |
| | v | 0.002* | 0.000 | 0.000 | 0.000 | 0.052 | 0.000* | 0.000 | 0.000 | 0.000 | 0.039 |
| Food Products (E &K) | i | 0.150 | **0.881** | **0.989** | 0.394 | 0.266** | **0.951** | **0.896** | **0.995** | 0.063 | 0.073** |
| | ii | 0.620 | 0.093 | 0.006 | 0.488 | 0.209 | 0.043 | 0.089 | 0.002 | **0.916** | 0.153 |
| | iii | 0.188 | 0.019 | 0.003 | 0.076 | 0.346 | 0.006 | 0.011 | 0.002 | 0.018 | 0.495 |
| | iv | 0.037 | 0.005 | 0.001 | 0.039 | 0.117 | 0.000 | 0.002 | 0.001 | 0.001 | 0.236 |
| | v | 0.005* | 0.003 | 0.000 | 0.003 | 0.062 | 0.000* | 0.002 | 0.000 | 0.001 | 0.043 |
| Non Food Products (F &L) | i | 0.093 | **0.734** | **0.990** | 0.377 | 0.309** | **0.827** | 0.762 | 0.994 | 0.082 | 0.078** |
| | ii | 0.619 | 0.187 | 0.010 | 0.494 | 0.138 | 0.132 | 0.198 | 0.006 | **0.874** | 0.131 |
| | iii | 0.221 | 0.059 | 0.000 | 0.076 | 0.343 | 0.033 | 0.029 | 0.000 | 0.027 | 0.514 |
| | iv | 0.065 | 0.019 | 0.000 | 0.052 | 0.170 | 0.009 | **0.009** | 0.000 | 0.017 | 0.239 |
| | v | 0.002* | 0.002 | 0.000 | 0.001 | 0.044 | 0.000* | 0.001 | 0.000 | 0.000 | 0.037 |
| Household durables (G & M) | i | 0.043 | 0.474 | **0.978** | 0.231 | 0.180** | 0.495 | 0.501 | **0.995** | 0.035 | 0.011** |
| | ii | 0.445 | 0.298 | 0.018 | 0.476 | 0.118 | 0.302 | 0.32 | 0.005 | 0.663 | 0.052 |
| | iii | 0.304 | 0.126 | 0.004 | 0.175 | 0.373 | 0.118 | 0.112 | 0.000 | 0.224 | 0.509 |
| | iv | 0.183 | 0.073 | 0.000 | 0.099 | 0.249 | 0.069 | 0.046 | 0.000 | 0.066 | 0.364 |
| | v | 0.025* | 0.029 | 0.000 | 0.018 | 0.079 | 0.018* | 0.020 | 0.000 | 0.012 | 0.063 |
| Housing (H & N) | i | 0.309 | **0.766** | **1.000** | 0.449 | 0.342** | **0.899** | 0.767 | 1.000 | 0.160 | 0.162** |
| | ii | 0.498 | 0.148 | 0.000 | 0.450 | 0.207 | 0.082 | 0.179 | 0.000 | **0.757** | 0.214 |
| | iii | 0.127 | 0.061 | 0.000 | 0.061 | 0.290 | 0.008 | 0.040 | 0.000 | 0.070 | 0.362 |
| | iv | 0.062 | 0.021 | 0.000 | 0.038 | 0.119 | 0.009 | 0.011 | 0.000 | 0.012 | 0.223 |
| | v | 0.004* | 0.004 | 0.000 | 0.001 | 0.042 | 0.002* | 0.003 | 0.000 | 0.001 | 0.038 |
| Services (I & J) | i | 0.122 | **0.711** | **1.000** | 0.376 | 0.254** | **0.793** | 0.77 | **1.000** | 0.059 | 0.062** |
| | ii | 0.566 | 0.192 | 0.000 | 0.462 | 0.155 | 0.119 | 0.161 | 0.000 | **0.854** | 0.175 |
| | iii | 0.215 | 0.054 | 0.000 | 0.086 | 0.378 | 0.062 | 0.042 | 0.000 | 0.051 | 0.469 |
| | iv | 0.092 | 0.037 | 0.000 | 0.075 | 0.182 | 0.026 | 0.024 | 0.000 | 0.034 | 0.261 |
| | v | 0.005* | 0.006 | 0.000 | 0.002 | 0.031 | 0.001* | 0.003 | 0.000 | 0.002 | 0.033 |

* (**) indicate Extreme false negative (positive) probability of all indicator variables.

The description of the different latent classes is as follows:

Class 1 (Rolling Stones)

About 9.4% of respondents were probabilistically assigned to this class. Around 71% of the respondents in this class expect that in general price increase similar to the current rate in the next 3 months but all respondents expect that in general price increase more than the current rate in next one year. For next one year, around (79 – 95) % of respondents expect that prices will increase for food products, non-food products, housing and services, respectively.

Class 2 (Simulators).

About 38.7% of respondents belong to this class. In this class, the respondents expects that the prices will increase more than the current rate in almost all the indicator variables i.e. in general (99% in next 3 months and 99% in next year); in food products (88% in next 3 months and 90 % in next one year); in non-food products (73% in next 3 months and 87% no change I prices in next one year); in housing (77% in next 3 months and 77% in next one year); and in services (71% in next 3 months and 77% in next one year). To household durables, the respondents were neutralizer in prices in next 3 months and in next one year.

Class 3 (Revenue Rocketers)

Around 28.2% of the respondents belong to this class. In this class, almost all the respondents are in favour of price increase more than the current rate in all the indicator variables.

Class 4 (Dormants/Neutralizers)

About 13.3% of the respondents belonging to this class. All the respondents in this class expect that the prices in general increase similar to current rate in next one year. In this class, the respondents expect that the prices will increase similar to the current rate for food products, non-food products, housing and services in the next one year, but for next 3 months the respondents are not clear.

Class 5 (Stagnants)

Around 10.3 % of the respondents belonging to this class, where the respondents are not clear about their expectations on different indicator variables.

Inconsistent responses occur when $x \neq d$. The probabilities of error in classification for the indicator variable D is given by $\pi_{d|x}^{D|X}$ for $d \neq x$. Extreme response biases will occur when we observe one of the following two scenarios:

(i) Case of extreme false positive probability = P(D= negative| X = improve); indicated by * in table 4.

(ii) Case of extreme false negative probability = P(D= positive| X = deteriorate); indicated by ** in table 4.

From table 4, it is noted that the respondents gave the extreme false negaive responses to the questions related to general (27%), food products (27%), non-food products (31%), household durables (18%), housing (34%) and services (25%), in next 3 months. For next one year, it seems that the respondents show their concern on housing by giving extreme false positive responses to the question on housing (16%).

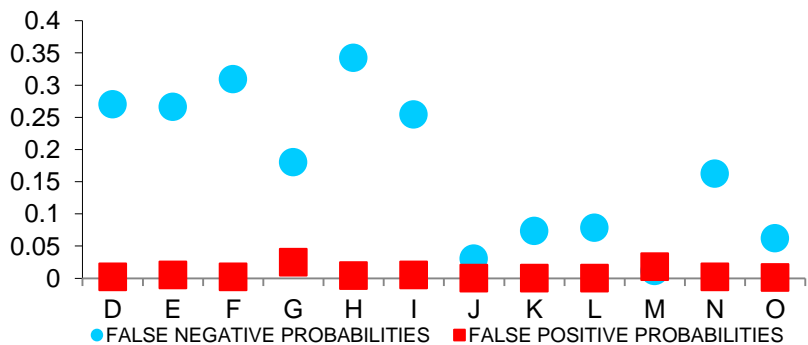

Figure 1: Extreme false negative and extreme false positive probabilities for different indicators.

Information about the extreme false positive and negative probabilities from Table 4 is summarized in Figure 1, where D, E, F, G, H and I indicates household expectations in next three months and J, K, L, M, N and O indicates the household expectations in next one year.

I have also calculated the probabilities of consistent classification for all indicators separately with the estimated class population shares. The probability is calculated as follows:

$$P_{JCC} = P(consistent\ classification\ for\ indicator\ vairable\ 'J'\ in\ all\ classes\ \forall J)$$

$$= P\begin{pmatrix} i \in n | i \in group\ '\sigma'\ by\ class\ population\ shares \\ and\ for\ indicator\ variable\ j\ \forall J;\ \sigma = 1,2,3 \end{pmatrix}$$

Similarly, the probability of misclassification for extreme cases is defined as

$$P^*_{JCC}$$

$$= P\begin{pmatrix} misclassification\ for\ all\ indicators\ with\ estimated\ class\ population \\ shares\ and\ for\ indicator\ variable\ J\ with\ \textbf{extreme}\ cases \end{pmatrix}$$

$$= P\begin{pmatrix} i \in n | i \in group\ '\sigma'\ by\ class\ population\ shares \\ and\ for\ extreme\ cases\ of\ all\ indicators \end{pmatrix} =$$

$$P\begin{pmatrix} i \in n | i \in group\ '\sigma'\ by\ class\ population\ shares \\ and\ for\ extreme\ cases\ of\ all\ indicators \end{pmatrix}.$$

A comparison of these two probabilities is important to determine whether the consumers' sentiment is being captured consistently through the indicators. The probabilities are given in Figure 2.

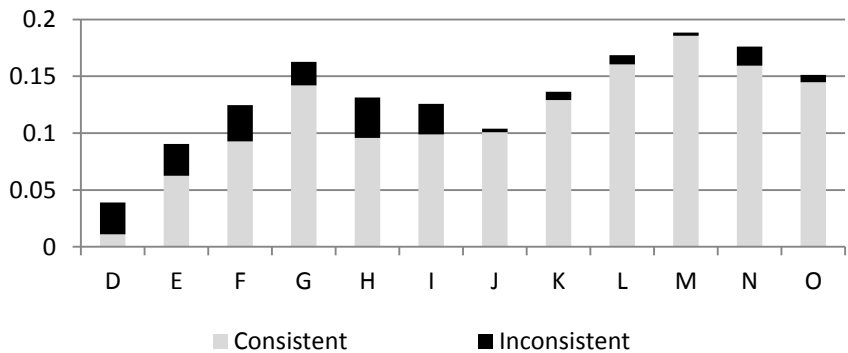

Figure 2: Consistent classification and misclassification probabilities of all indicators

The height of the grey bar gives the probability of consistent classification of each indicator while the height of the black bar gives the probability of inconsistent classification for that indicator. It is noted that the probability of consistent classification of all indicators is much higher than the inconsistent classification except in case of general household's expectation on price in next 3 months, where probability of misclassification is higher than the classification. On average, the probability of consistent misclassification is 0.0179 and consistent classification is 0.1152.

**5. Conclusion**

The main role of the household inflation expectation survey is the development of improved survey measures of inflation expectations. It is believe that the new survey measures will enhance the ability of policymakers to monitor key aspects of consumer inflation expectations much remains to be learned about how consumers form and act on their expectations.

In many countries expectation surveys which means to collect information from consumers about their expectations on prices. Unfortunately, before doing so, the reliability of responses is often not taken into account, which leads to biases creeping in and affecting the reliability of the indices. In particular extreme response bias is an important category of bias that may distort the results of the survey. With latent class analysis, approximately 34% of the respondents are complaining against the government by giving the response bias on different indicator variables. Respondent's shows more concern in general, the prices increases more than the current rate by giving extreme responses on housing in next quarter. Further, it would be interesting to study such a scenario with the help of a panel data where we would be able to identify the impact of time on changes in consumer expectations on inflation.

**Appendix I:** Set of questions from the standardized Inflation expectations survey of households Questionnaire which is used in our analysis.

Block 2: Expectations of respondent on prices in next 3 months

| Options | General (D) | Food Products (E) | Non- food products (F) | Household durables (G) | Housing (H) | Services (I) |
|---|---|---|---|---|---|---|
| i. Price increase more than current rate | | | | | | |
| ii. Price increase similar to current rate | | | | | | |
| iii. Price increase less than current rate | | | | | | |
| iv. No change in prices. | | | | | | |
| v. Decline in prices. | | | | | | |

Block 3: Expectations of respondent on prices in next one year

| Options | General (J) | Food Products (K) | Non- food products (L) | Household durables (M) | Housing (N) | Services (O) |
|---|---|---|---|---|---|---|
| vi. Price increase more than current rate | | | | | | |
| vii. Price increase similar to current rate | | | | | | |
| viii. Price increase less than current rate | | | | | | |
| ix. No change in prices. | | | | | | |
| x. Decline in prices. | | | | | | |

Source: IESH, conducted by Reserve Bank of India.